\documentclass[conference,9pt]{IEEEtran}
\IEEEoverridecommandlockouts
\usepackage{cite}
\usepackage{caption}
\usepackage{amsmath,amssymb,amsfonts}
\usepackage{algpseudocode}            
\usepackage{algorithm}
\usepackage{graphicx}
\usepackage{textcomp}
\usepackage{xcolor}
\usepackage{multirow}
\usepackage{paralist}
\usepackage{amsmath}
\usepackage{subfig}

\usepackage{float}
\usepackage{adjustbox}
\def\BibTeX{{\rm B\kern-.05em{\sc i\kern-.025em b}\kern-.08em
    T\kern-.1667em\lower.7ex\hbox{E}\kern-.125emX}}

\begin{document}

\title{\LARGE KiD: A Hardware Design Framework Targeting Unified NTT Multiplication for CRYSTALS-Kyber and CRYSTALS-Dilithium on FPGA\vspace{-0.1 in}
\thanks{Support for this work was provided by C3i (cybersecurity and cybersecurity for Cyber-Physical Systems) Innovation Hub, IIT Kanpur.}
}

\author{\IEEEauthorblockN{Suraj Mandal, Debapriya Basu Roy}
\IEEEauthorblockA{\textit{Department of Computer Science \& Engineering},
\textit{Indian Institute of Technology Kanpur, India}\\
\{surajmandal,dbroy\}@cse.iitk.ac.in}
}
\maketitle
\begin{abstract}
Large-degree polynomial multiplication is an integral component of post-quantum secure lattice-based cryptographic algorithms like CRYSTALS-Kyber and Dilithium. The computational complexity of large-degree polynomial multiplication can be reduced significantly through Number Theoretic Transformation (NTT). In this paper, we aim to develop a unified and shared NTT architecture that can support polynomial multiplication for both CRYSTALS-Kyber and Dilithium. More specifically, in this paper, we have proposed three different unified architectures for NTT multiplication in CRYSTALS-Kyber and Dilithium with varying number of configurable radix-2 butterfly units. Additionally, the developed implementation is coupled with a conflict-free memory mapping scheme that allows the architecture to be fully pipelined. We have validated our implementation on Artix-7, Zynq-7000 and Zynq Ultrascale+ FPGAs. Our standalone implementations for NTT multiplication for CRYSTALS-Kyber and Dilithium perform better than the existing works, and our unified architecture shows excellent area and timing performance compared to both standalone and existing unified implementations. This architecture can potentially be used for compact and efficient implementation for CRYSTALS-Kyber and Dilithium.


\end{abstract}
\begin{IEEEkeywords}
Kyber, Dilithium, PQC, NTT, FPGA
\end{IEEEkeywords}
\section{Introduction}
Recent growth in quantum computers has catalyzed the process of replacing classical public key cryptography with quantum secure algorithms. NIST has selected CRYSTALS-Kyber as the key encapsulation mechanism (KEM) in its round 4, and CRYSTALS-Dilithium is selected as one of the candidates for the digital signature algorithm. CRYSTALS-Kyber and CRYSTALS-Dilithium are lattice-based cryptographic algorithms based on \emph{learning with error} problem. Execution of these two algorithms requires large polynomial multiplication, and classical polynomial multiplication with $\mathcal{O}{(n^2)}$ complexity turns out to be inefficient for this purpose.  Number Theoretic Transformation (NTT) based polynomial multiplication, on the other hand, has the asymptotic complexity of $\mathcal{O}{(n\log n)}$ and therefore is more suitable for large degree polynomial multiplication. 
In this paper, we aim to develop a unified NTT-based polynomial multiplier that can support both CRYSTALS-Kyber and Dilithium. As NIST has already zeroed on CRYSTALS-Kyber and Dilithium as the candidates for KEM and digital signature, our proposed unified architecture would help to develop lightweight and compact architectures for post-quantum secure public key infrastructure.

CRYSTALS-KYBER and CRYSTALS-Dilithium are defined over quotient ring ${Z}_q[x]/(X^n +1)$ with $q=3329$ and $8380417$ respectively. Thus, the coefficients of polynomials defined for Kyber and Dilithium are $12$ and $23$ bits, respectively. 
Therefore, the space required to store two coefficients of Kyber's polynomial is nearly identical to that of a single coefficient of Dilithium's polynomial. This feature motivated us
to develop a unified architecture for NTT multiplication, supporting both Kyber and Dilithium. 
We have explored the design space of radix-2-based unified NTT polynomial multiplier and have proposed three designs. 
Our unified NTT architecture for CRYSTALS-Kyber/CRYSTALS-Dilithium outperforms the existing work~\cite{aikata2022kali} that focuses on the same objective. We also have shown that even our results for standalone designs of NTT multiplication for Kyber and Dilithium achieve better Area Dealy Product (ADP) than the existing designs in the literature. The contributions of this work are summarized below:
\begin{compactitem}
    \item We have designed a standalone, fully pipelined, and scalable radix-2 Kyber/Dilithium NTT multiplication unit with a lesser area overhead and a significant improvement over frequency. Using these as our base modules, we have proposed three unified NTT multiplication units for Kyber and Dilithium:
    \begin{compactitem}
        \item \textbf{Design 1:} The first design is based on $2$ radix-2 BFUs for Kyber that can also be used as $1$ radix-2 BFU of Dilithium.
        \item \textbf{Design 2:} The second design is based on $4$ radix-2 BFUs of Kyber that can be used as $2$ radix-2 BFUs of Dilithium.
        \item \textbf{Design 3:} Our third design is based on $8$ radix-2 BFUs of Kyber that can be used as $4$ radix-2 BFUs of Dilithium.
    \end{compactitem}

\item We have proposed an efficient conflict-free memory mapping scheme that allows us to pipeline the developed BFUs without any stalls. This memory mapping scheme targets BRAMs on FPGA with \emph{simple dual port} configuration. 
    \item We have implemented and verified our proposed configurable radix-2-based NTT multiplication units on Artix-7 and Zynq Ultrascale FPGAs. The result shows that our proposed architecture exhibits superior ADP (Area-Delay Product) values compared to existing implementations.

\end{compactitem}
The rest of the paper is organized as follows. In section \ref{sec:relwork}, we have summarized the previous standalone and unified NTT-based polynomial multiplication implementation for Kyber/Dilithum. Section \ref{sec:background1} describes the required background, and section \ref{DesignChoice} focuses on the NTT multiplication architecture along with the conflict-free memory mapping scheme. Our proposed unified NTT multiplication unit has been described in section \ref{sec:unified}. Section \ref{sec:experiment} provides the comparative analysis of our proposed implementation and finally, in section~\ref{sec:conc}, we conclude the paper.
\section{Related Works}\label{sec:relwork}
This section summarizes the existing works that focus on the implementations of NTT architecture for Kyber and Dilithium.
Authors in \cite{xing2021compact} proposed a hardware accelerator for CRYSTALS-Kyber with pipelined accesses of coefficients for NTT/inverse NTT (NTT$^{-1}$)/ pointwise multiplication (PWM) with an overhead of $1737$ look-up tables (LUTs), $2$ DSP blocks, $3$ Block RAMs (BRAMs) and a latency of 448 clock cycles. 
Later, authors in \cite{bisheh2021high} proposed a hardware-friendly NTT polynomial unit with a lightweight modular reduction for CRYSTALS-Kyber. The design has a latency of $324$ clock cycles with an overhead of $801$ look-up tables (LUTs), $4$ DSPs, and $2$ BRAMs with a frequency of $222$ MHz. Authors in \cite{zhang2021towards} proposed a low area NTT multiplication unit for Kyber with pipeline stalls of $6$ clock cycles after each stage during NTT/NTT$^{-1}$. The work of~\cite{dang2022high} proposed fully pipelined architecture for NTT polynomial multiplication for CRYSTALS-Kyber, requiring only $2$ DSPs and $2$ BRAMs. In \cite{lil2022high}, authors have implemented a lightweight modular reduction unit for signed numbers in Kyber. The recent work~\cite{li2022reconfigurable} proposes an NTT multiplication unit for Kyber. However, the modular addition/subtraction in~\cite{li2022reconfigurable} will lead to faulty values.

The NTT polynomial unit for CRYSTALS-Dilithium in \cite{beckwith2021high} utilizes a LUT overhead of $4509*2$ for $4$ BFUs with $16$ DSPs and incorporates a pipelined read-write accesses. Later on, authors in \cite{zhao2022compact} designed an NTT multiplication unit for Dilithium with improved area overhead that consumes $1919$ LUTs, $8$ DSPs with additional $893$ LUTs and $2$ DSPs for head-tail reordering. Authors in \cite{wang2022efficient} proposed a radix-4 based fully pipelined NTT multiplication unit for Dilithium that exhibits a significantly improved area overhead with $2386$ LUTs and $8$ DSPs. Authors in \cite{wang2022efficient} have used a Solinas modular reduction unit with $260$ LUTs and $2$ DSPs. In \cite{gupta2023lightweight}, authors proposed a hardware accelerator for Dilithium, including an NTT multiplication unit with an overhead of $2759$ LUTs $4$ DSP and $7$ BRAMs. 

Although we can find a sufficient number of works that implement NTT polynomial multiplication units for Kyber and Dilithium, the unified NTT architecture design for Kyber and Dilithium is still unexplored. 
Authors in \cite{aikata2022kali} proposed a unified architecture ($4$ BFUs of Kyber with reconfigurability of two BFUs for Dilithium) with $3487$ LUTs, and $4$ DSPs and a frequency of $270$ MHz.
 \begin{figure}[h]
    \centering
    \includegraphics[scale=0.5]{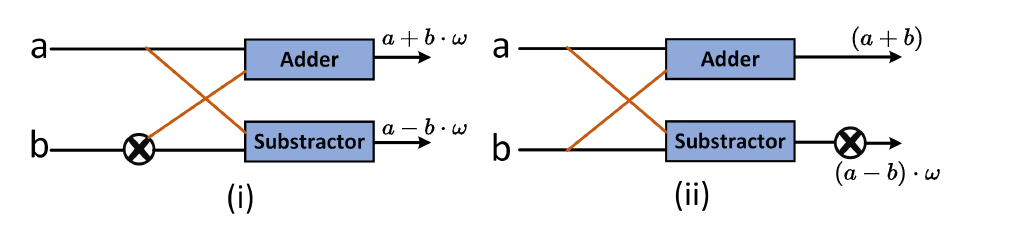}
    \caption{(i)Cooley-Tukey and (ii)Gentleman-Sande Butterfly Units}
    \label{fig:bfuconstruction}
    \vspace{-0.1 in}
\end{figure}
\section{Background}
\label{sec:background1}
Polynomial Multiplication of CRYSTALS-Kyber and CRYSTALS-Dilithium is performed on a $n$-degree polynomial in the ring ${Z}_q[x]/(X^n +1)$, referred to as negative wrapped convolution where $q$ is modulus over the ring. NTT multiplication of two polynomials $A(x) = \sum_{i=0}^{n-1} a_{i}x^{i}$ and $B(x) = \sum_{i=0}^{n-1} b_{i}x^{i}$ ($a_{i}, b_{i} \in \mathbb{Z}_q$), can be defined as $NTT^{-1}(NTT(A) \bullet NTT(B))$, where $\bullet$ signifies the pointwise multiplication (PWM). The equation for NTT and NTT$^{-1}$ on polynomial $A$ for negative wrapped convolution is given by Eq.~\eqref{eq:eqntt}, $\gamma$ is $2n^{th}$ root of unity i.e. $\gamma^{2n} \equiv 1 \mod q$ and  $\omega$ is the $n^{th}$ root of unity with $\gamma= \sqrt{\omega}$. Different powers of $\gamma$ and $\omega$, known as twiddle factors, are accessed as shown in Eq.\eqref{eq:eqntt}.
\begin{equation} \label{eq:eqntt}
\footnotesize
\Tilde{A_j} = \sum_{i=0}^{n-1} {\gamma^i \cdot \omega_{n}}^{ij} \cdot A_i, 
A_j =\frac{1}{n} \cdot \gamma^{-j} \cdot \sum_{i=0}^{n-1} {\omega_{n}}^{-ij} \cdot \Tilde{A_j}
\end{equation}
\paragraph{NTT in CRYSTALS-Kyber}
\label{kyber}
Polynomial multiplication in Kyber is equivalent to negative wrapped convolution with modulus $q=3329$ and $n=256$. But $512^{th}$ primitive root of unity does not exist for Kyber; rather, Kyber has $256^{th}$ root of unity. So, Kyber performs incomplete NTT i.e. NTT of odd and even coefficients are computed independently~\cite{avanzicrystals}. To process the coefficients during NTT/NTT$^{-1}$, Cooley-Tukey and Gentleman-Sande butterfly unit (BFU) constructions are used as shown in Fig.~\ref{fig:bfuconstruction} (a and b are two coefficients and $\omega$ is one of the twiddle factors)~\cite{poppelmann2015high}. These BFUs will be the basic building blocks for our proposed implementation. Point-wise multiplication (PWM) in the case of Kyber involves $5$ modular multiplications and $2$ modular additions in $Z_q$. Point-wise multiplication of two NTT transformed polynomials $\Tilde{A}$ and $\Tilde{B}$ is shown in Eq.~\eqref{eq:pwmkyber} where $\psi_i$ are twiddle factors.
\begin{equation} \label{eq:pwmkyber}
\footnotesize
    res_0 = \Tilde{A}_{2i} \cdot \Tilde{B}_{2i} + \Tilde{A}_{2i+1} \cdot \Tilde{B}_{2i+1}\cdot {\psi_i},
    res_1 = \Tilde{A}_{2i} \cdot \Tilde{B}_{2i+1} + \Tilde{A}_{2i+1} \cdot \Tilde{B}_{2i}
\end{equation}
\paragraph{NTT in CRYSTALS-Dilithium}
Polynomial multiplication in CRYSTALS-Dilithium is also performed in the ring ${Z}_q[x]/(X^n +1)$ where $q=8380417$ and $n=256$. In this case, $512^{th}$ root of unity exists, and we can deploy complete NTT for Dilithium.
 After the transformation of the coefficients into the NTT domain, pointwise multiplication (PWM) is performed by multiplying each coefficient of one polynomial with each coefficient of another polynomial without any additional overhead. BFU construction for processing the coefficients in Dilithium is also the same as Kyber.
 \vspace{-0.05 in}
\section{Design Choices of the Proposed Framework}
\begin{figure}[]
    \centering
    \includegraphics[clip,scale=0.3]{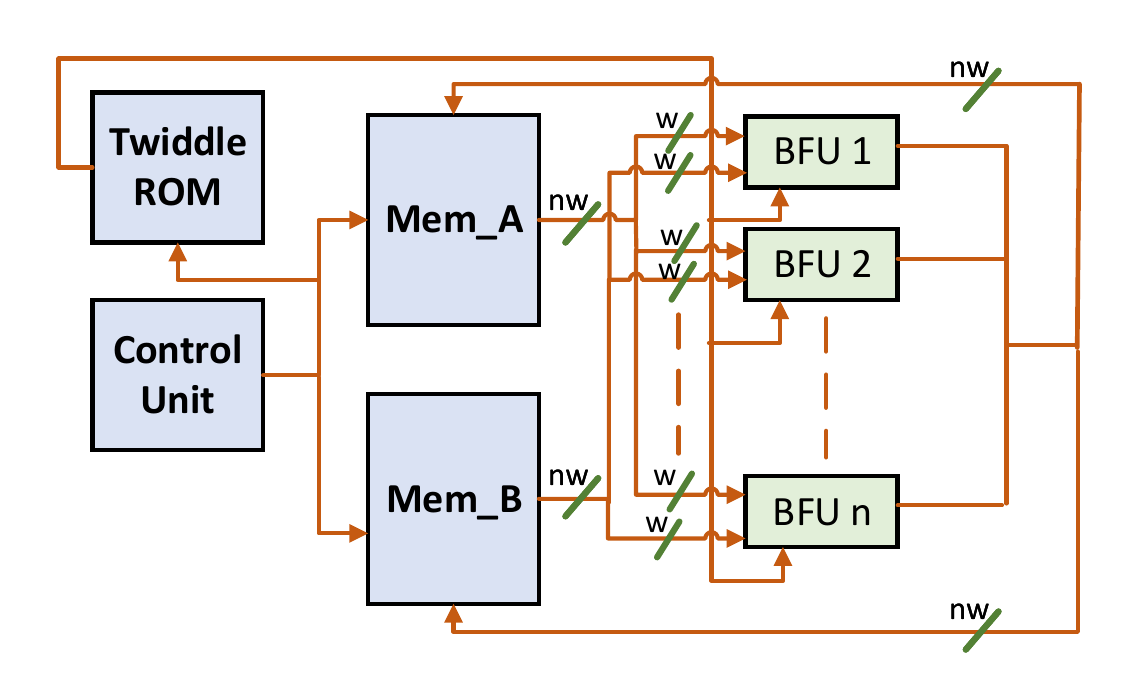}
    \caption{Design Choice for NTT Multiplication Unit}
    \label{fig:designChoice}
    \vspace{-0.2 in}
\end{figure}

\vspace{-0.05 in}
\label{DesignChoice}
We have shown the generic design framework of our proposed implementation in Fig.~\ref{fig:designChoice}.
To design the radix-2-based scalable NTT multiplication core architecture, we have stored the input polynomial coefficients into two simple dual-port Block RAMs (BRAMs) referred to as $Mem_A$ and $Mem_B$ in Fig.~\ref{fig:designChoice}. The width and the depth of the BRAMs are determined according to the parallel BFUs used during computation. For a $n$ degree polynomial, if $t$ number of BFUs are used in parallel, then the size of each BRAM should be $(t*$ size of coefficient)$* d$ where $d=\frac{n}{t*2}$ is the depth of the two BRAMs. For example, to use $2$ BFUs in parallel for NTT operation in CRYSTALS-Kyber, the width of each BRAM should be $(2*12)$ ($t=2$ and size of coefficients= $12$ bits), and the depth of each memory will be $\frac{256}{2*2}=64$ ($n=256$). The simple dual-port BRAMs allow parallel read and write operations through their dual ports, allowing us to pipeline the architecture without any hazards.
In addition to two simple dual-port RAMs and the BFUs, our NTT polynomial multiplication unit consists of the three following components:
\begin{compactitem}
    \item A single port ROM for storing the twiddle factors.
    \item A single port ROM for accessing the read/write addresses with additional flag bits.
    \item A control counter that includes a memory address generator and a twiddle ROM counter.
\end{compactitem}
To reduce the look-up table (LUT) consumption, instead of generating addresses from the control unit, we have hard-coded the address values for $Mem_A$ and $Mem_B$ inside a BRAM configured as a single port ROM. The memory also includes additional flag bits that decide the read/write order of the coefficients from $Mem_A$ or $Mem_B$. 

\subsection{Scalable and Conflict Free Memory Mapping}
\label{sec:conflictfree}
 We have used Cooley-Tukey-based decimation in time (DIT) NTT and Gentleman-Sande-based decimation in frequency (DIF) NTT$^{-1}$~\cite{poppelmann2015high}.
Cooley-Tukey-based DIT NTT algorithm processes input coefficients in standard order and produces output in bit-reverse order, whereas Gentleman-Sande-based DIF NTT$^{-1}$ consumes inputs in bit-reverse order and produces output in standard order.

 NTT/NTT$^{-1}$ of a $n$-degree polynomial is performed in $\log_2 n$ stages \cite{poppelmann2015high}. Before reading the coefficients from a memory address in the next stage, we must ensure that the write operation on that address from the previous stage has been completed. To overcome this memory-conflict issue, authors in~\cite{chen2022cfntt} proposed a conflict-free memory mapping scheme without any pipeline stalls for radix-$2^k$ NTT. This method of conflict-free memory mapping needs a higher number of BRAMs with an increasing number of BFUs in parallel with interleaved memory access banks. Authors in \cite{dang2022high,aikata2022kali} have also used a fully pipelined NTT core implementation. But no algorithms in \cite{dang2022high,aikata2022kali} have been provided for the fully pipelined memory mapping scheme.
 \begin{figure}[h]
    \centering
    \includegraphics[clip,scale=0.3]{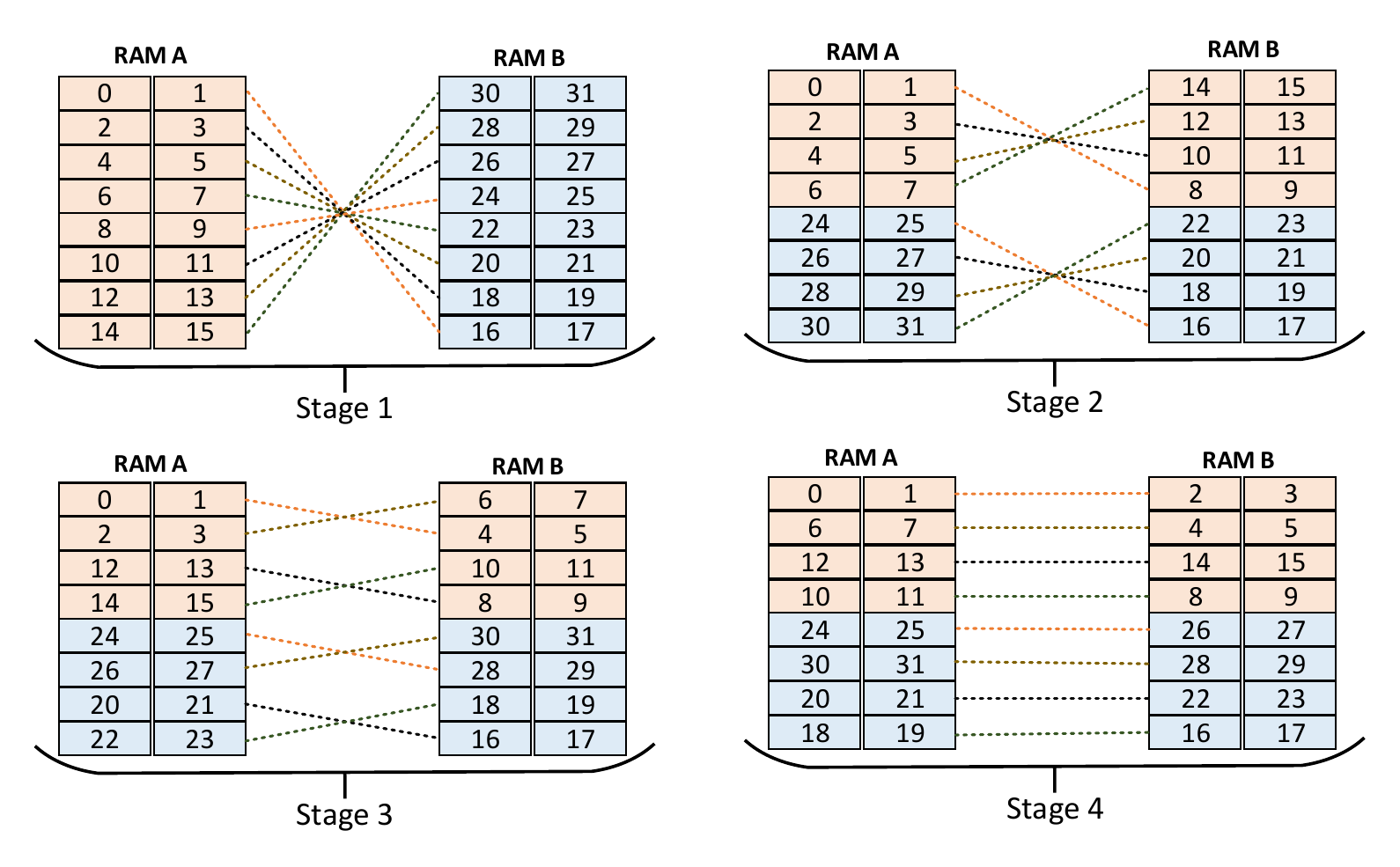}
    \caption{Memory Read-Write Access During NTT.}
    \label{fig:memacess}
    \vspace{-0.15 in}
\end{figure}
For maintaining the pipelined access of the coefficients, we have opted for a simple and efficient way to perform radix-2 NTT/NTT$^{-1}$ operations using two simple dual-port RAMs as given in Algorithm \ref{algorithm1}. Authors in \cite{chen2022cfntt} have used $2.5$ and $5$ and $10$ BRAMS for $2$, $4$ and $8$ BFU configurations for radix-2 BFUs, respectively. Whereas our memory mapping scheme consumes $2$, $4.5$ and $5.5$ for the same configurations. This shows that our proposed memory mapping scheme is superior to~\cite {chen2022cfntt} for parallel radix-2 configurations.
\begin{algorithm}[]
\scriptsize
\caption{Pipelined Memory Addressing for NTT/NTT$^{-1}$}
\label{algorithm1}
    \begin{algorithmic}[1]
        \Procedure{Addressing}{ $ ch, d$}
        \If{$ch$==1}
            \State $start=1$, $end=\log_2d$
        \Else
             \State $start=\log_2d$, $end=1$
        \EndIf
            \For {{i=start to end}}
            \State $m=2^i$, $p_2=\frac{m}{2}$
                 \For{{j=0 to $N$ by $p_2$}}
                    \State $k_1=j$, $k_2=p_2-1+j$
                    \For{{k=0 to $p_2$}}
                        \If{$k$ is even}
                            \State $address_A=k_1$, $address_B=k_2$
                        \ElsIf{$k$ is odd}
                            \State $address_A=k_2$, $address_B=k_1$
                            \State $k_1=k_1+1$, $k_2=k_2-1$
                        \EndIf
                    \EndFor
                 \EndFor
            \EndFor
        \State return ($address_A$, $address_B$)
        \EndProcedure
    \end{algorithmic}
\end{algorithm}

The in-order input sequence of coefficient storage in $Mem_A$ and $Mem_B$ for a $32$-degree polynomial is given in stage 1 of Fig.~\ref{fig:memacess}. Our algorithm (Algorithm. \ref{algorithm1}) takes the start and end index as inputs and produces two addresses i.e. $address_A$ and $address_B$ as read/write indexes for $Mem_A$ and $Mem_B$. The parameter $ch$ when $0$, the Algorithm~\ref{algorithm1} generates the addresses for NTT. When this parameter $ch=1$, this algorithm produces the address for NTT$^{-1}$. 
According to Algorithm \ref{algorithm1}, an example of the data flow between the RAMs in each stage for a radix-2 NTT with a 32-degree polynomial using two parallel BFUs is given in Fig.~\ref{fig:memacess}.  For a memory depth of $d$, the read/write address sequence for 1st stages of BFU operations will be $(\{\{0,d\},\{d,0\},\{1,d-1\},\{d-1,1\},...,\{\frac{d}{2}-1,\frac{d}{2}\}, \{\frac{d}{2},\frac{d}{2}-1\}\})$. For the second stage operations, the read/write address sequence will be $(\{\{0,\frac{d}{2}-1\},\{\frac{d}{2}-1,0\},\{1,\frac{d}{2}-2\},\{\frac{d}{2}-2,1\},...,\{\frac{d}{2}+1,d\},\{d,\frac{d}{2}+1\},...\})$. For third stages the read/write addresses will start from $\frac{d}{4}$ and so on. According to the algorithm~\ref{algorithm1}, a read-write conflict may occur during the end of the first-stage operations and the beginning of the second-stage operations. To overcome this problem, we have reversed the memory access of the last half of the address sequence for 1st stage of NTT. 
The condition to maintain fully pipelined implementation of radix-2 NTT is shown in Eq.~\eqref{eq:eqpipeline}, and our conflict-free memory mapping scheme satisfies this condition. 
\begin{equation}\label{eq:eqpipeline}\footnotesize
    Pipeline\ Depth \leq \frac{Depth\ of\ Coefficient\ Memory}{2}.
\end{equation}
So, to maintain the conflict-free accesses, the maximum pipeline depth for a $256$ degree polynomial with $2$, $4$ and $8$ BFU accesses in parallel should be at most $\frac{256}{2*4}=32$, $\frac{256}{2*8}=16$ and $\frac{256}{2*16}=8$ respectively. If we increase the parallel BFUs to $16$, the maximum pipeline depth will be $\frac{256}{2*32}=4$, which would increase the critical path of the design significantly. 
\begin{figure}[b]
    \centering
    \includegraphics[clip,scale=0.23]{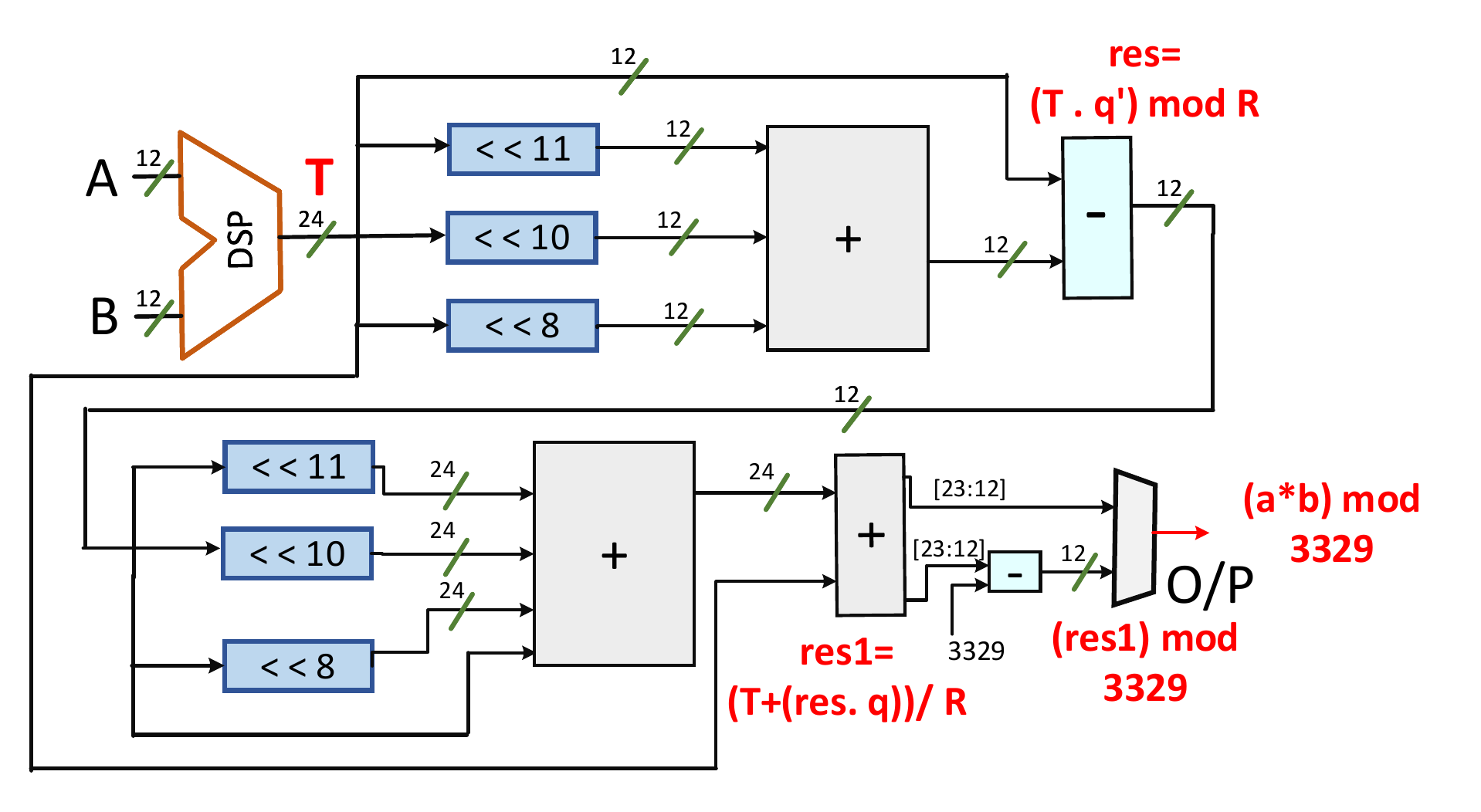}
    \caption{Montgomery Modular Multiplier for Kyber}
    \label{fig:montkyber}
\end{figure}
\vspace{-0.05 in}
\subsection{Polynomial Multiplication Unit for Kyber}
\begin{figure}[h]
    \centering
    \subfloat[\centering Modular Adder]{{\includegraphics[width=0.2\textwidth]{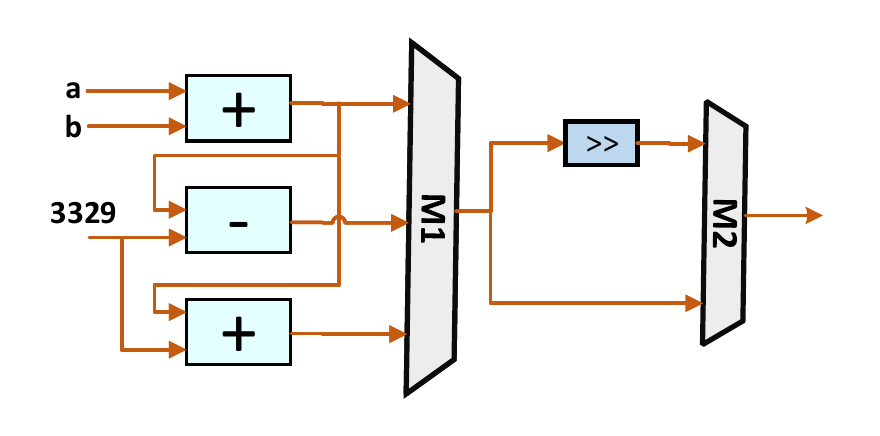} }}%
    \qquad
    \subfloat[\centering Modular Subtractor]{{\includegraphics[width=0.2\textwidth]{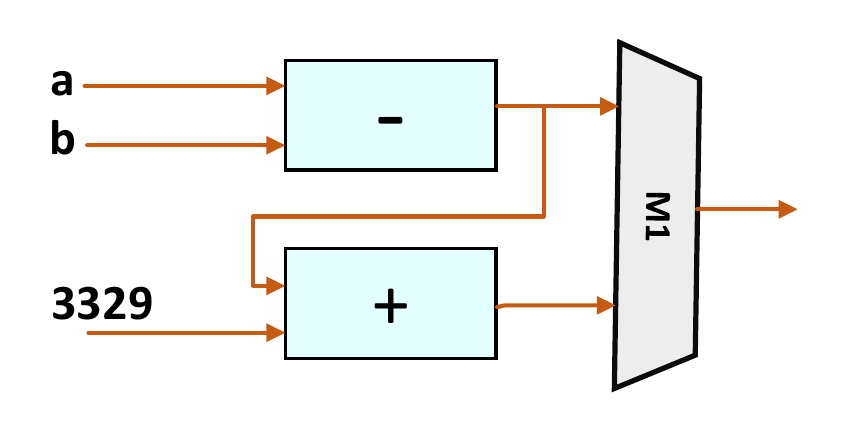} }}%
    \caption{(a)Modular Adder for Kyber. (b) Modular Subtractor for Kyber.}%
    \label{fig:addsubmethod}%
    \vspace{-0.2 in}
\end{figure}

\vspace{-0.05 in}
\label{polykyber}
\paragraph{Coefficient Multiplier}
CRYSTALS-Kyber has a 12-bit modulus $3329$ with a structure $2^{11}+2^{10}+2^8+1$. 
Our implementation of coefficient multiplication is based on Montgomery modular multiplication~\cite{montgomery1985modular}. Montgomery multiplication requires $3$ multiplications. Apart from the standard multiplication between two coefficients, the other two multiplications are done with $q$ and $q'$, where $q.q'= -1 \ \text{mod} \ R$ with $R=2^{12}$. The values of $q$ and $q'$ are $3329$ and $3327$, respectively and we perform these multiplications using simple shifters and adders. The overhead of our developed coefficient multiplication are $71$ LUTs and $1$ DSP for one $12 \times 12$ bit multiplication. An architectural diagram of our modular Montgomery multiplication unit for Kyber is shown in Fig.\ref{fig:montkyber}. 
\paragraph{Modular Adder and Subtractor}
The modular adder for Kyber is depicted in Fig.~\ref{fig:addsubmethod}.(a) where $a,b$ are two coefficients that will be added. During NTT operation, modular addition of $a$ and $b$ can be defined by $a+b-3329$ or $a+b$ depending upon the condition if $a+b>3329$ or not. However, during NTT$^{-1}$ operation, in each stage, the result of the modular adder needs be divided by $2$. This is done to avoid an extra division by $n$, required for NTT$^{-1}$ operation as shown in Eq.~\eqref{eq:eqntt}~\cite{poppelmann2015high}. The operation $2^{-1}.(a+b)$ can be simply done by one right shift if $a+b$ is even. However, if $a+b$ is odd, then we need to compute $a+b+3329$ and then perform the right shift.
To perform an exact division by 2, we have configured MUX M1 of Fig.~\ref{fig:addsubmethod} adder by following four conditions:
\begin{compactitem}
    \item \textbf{$\mathbf{a+b>3329}$ and $\mathbf{a+b}$ is odd}: This means that after computing $a+b-3329$ the answer would be even, hence we would produce $a+b-3329$ as output of MUX M1. 
    \item \textbf{$\mathbf{a+b>3329}$ and $\mathbf{a+b}$ is even}: This implies that after computing $a+b-3329$ the answer would be odd; hence, for right-shift we need to add $3329$ again to this value. Therefore, the final output from MUX M1 would be $a+b$,   

    \item \textbf{$\mathbf{a+b<3329}$ and $\mathbf{a+b}$ is odd}: In this case, the output of the MUX M1 would be $a+b+3329$.
    \item \textbf{$\mathbf{a+b<3329}$ and $\mathbf{a+b}$ is even}: In this case, the output of the MUX M1 would be $a+b$.
\end{compactitem}
For MUX M2 in Fig. \ref{fig:addsubmethod}, the right shift by $1$ (division by 2) operation is selected during the NTT$^{-1}$ operation. Modular subtractor (Fig. \ref{fig:addsubmethod}.(b)) has been configured quite simply with $a-b+3329$ or $a-b$ depending upon the most significant bit of $a-b$. For NTT$^{-1}$, we need to compute $\frac{\omega}{2}(a-b)$. This can be done by multiplying each twiddle factor with $2^{-1}$, storing them in the \emph{twiddle ROM} and using them during multiplication.

The authors in \cite{li2022reconfigurable} reported a lightweight modular adder/subtractor unit for Kyber that also performs multiplication with $2^{-1}$ in each stage of NTT$^{-1}$. During modular addition, this work performs $a+b-3329$ only when the $a+b$ is a $13$ bit value (section III.B of~\cite{li2022reconfigurable}). However, this will give an erroneous result as it is possible for $a+b$ to be greater than $3329$ and still be a $12$ bit value.

 The NTT multiplication unit for Kyber uses 4-coefficients simultaneously with 2 BFUs. To fit a $256$-degree polynomial and to support our pipelined memory mapping scheme, we have used two simple dual-port RAM of size $128*24$. One $18K$ single port ROM has been used to store the twiddle factors, and another $18K$ single port ROM (address ROM) is used to embed the control unit. 
\begin{figure}[]
    \centering
    \includegraphics[clip,scale=0.2]{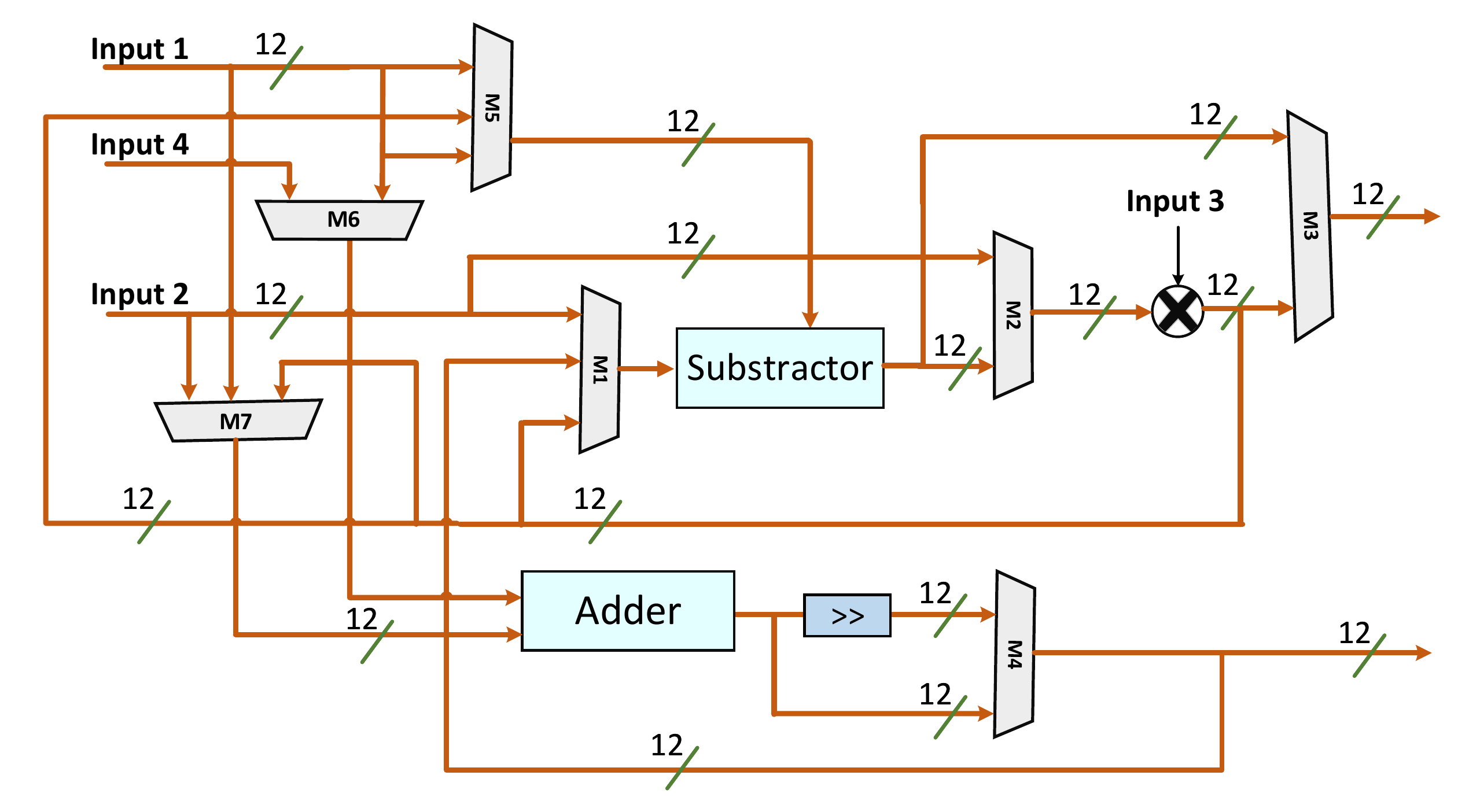}
    \caption{BFU Unit for Kyber}
    \label{fig:bfukyber2}
    \vspace{-0.2 in}
\end{figure}
The butterfly unit of Kyber has been configured to act as a BFU for NTT/NTT$^{-1}$/PWM operations depending upon the select lines of multiplexers placed in different positions of the data path. We have used two different BFU architectures for Kyber to handle the complex pointwise multiplication as mentioned in section \ref{sec:background1}. The operation can be further simplified with $4$ multiplications instead of $5$ using the Karatsuba method~\cite{xing2021compact}. For PWM, we have followed the work of~\cite{xing2021compact} and have generated two stages, $PWM0$ and $PWM1$, in a pipelined fashion to complete the PWM execution in $256$ clock cycles. The architecture of BFU of CRYSTALS-Kyber is given in Fig.~\ref{fig:bfukyber2}. In this figure (Fig.~\ref{fig:bfukyber2}), input 1 and input 2, input 3 and input 4 are coming from BRAMs. Input 1 , input 2 and input 3 are used during NTT/NTT$^{-1}$ whereas input 4 is only used during PWM operations. Input 3 is used as twiddle input during NTT/NTT$^{-1}$ and the coefficient inputs during PWM operations.
One BFU of Kyber does not include multiplexer $M6$ and operates with $6$ control signals, whereas another BFU uses $M6$ to perform with $12$ control signals. The configuration of the control MUXes during NTT/INTT/PWM operations for both the BFUs are given in Table.~\ref{tab:kyberbfucontrol}.
 Our pipelined NTT polynomial architecture has a latency of $448$ cycles for NTT/NTT$^{-1}$ and $256$ cycles for PWM with a  pipeline depth of $11$. The NTT polynomial unit for Kyber with 2 BFU has a latency of $1152$ with an overhead of $698$ LUTs, $865$ FFs, $2$ DSP s, $2$ BRAMs and achieved a frequency of $500$ MHz on Zynq Ultrascale+ FPGA.
 \begin{table}[h]
\centering
\scriptsize
\begin{tabular}{|c|c|cccccc|c|c|c|c|c|c|}
\hline
                       &      & \multicolumn{1}{c|}{\textbf{11}} & \multicolumn{1}{c|}{\textbf{10}} & \multicolumn{1}{c|}{\textbf{9}} & \multicolumn{1}{c|}{\textbf{8}} & \multicolumn{1}{c|}{\textbf{7}} & \textbf{6} & \textbf{5} & \textbf{4} & \textbf{3} & \textbf{2} & \textbf{1} & \textbf{0} \\ \hline
\multirow{4}{*}{BFU 1} & NTT  & \multicolumn{1}{c|}{0}           & \multicolumn{1}{c|}{0}           & \multicolumn{1}{c|}{0}          & \multicolumn{1}{c|}{0}          & \multicolumn{1}{c|}{0}          & 0          & 0          & 0          & 1          & 0          & 0          & 1          \\ \cline{2-14} 
                       & INTT & \multicolumn{1}{c|}{0}           & \multicolumn{1}{c|}{0}           & \multicolumn{1}{c|}{1}          & \multicolumn{1}{c|}{0}          & \multicolumn{1}{c|}{1}          & 1          & 1          & 1          & 0          & 1          & 0          & 0          \\ \cline{2-14} 
                       & PWM0 & \multicolumn{1}{c|}{1}           & \multicolumn{1}{c|}{1}           & \multicolumn{1}{c|}{0}          & \multicolumn{1}{c|}{1}          & \multicolumn{1}{c|}{0}          & 0          & 0          & 0          & 1          & 0          & 1          & 0          \\ \cline{2-14} 
                       & PWM1 & \multicolumn{1}{c|}{0}           & \multicolumn{1}{c|}{0}           & \multicolumn{1}{c|}{0}          & \multicolumn{1}{c|}{0}          & \multicolumn{1}{c|}{1}          & 0          & 0          & 1          & 1          & 0          & 0          & 0          \\ \hline
\multirow{4}{*}{BFU 2} & NTT  & \multicolumn{6}{c|}{\multirow{4}{*}{x}}                                                                                                                                                & 0          & 0          & 0          & 1          & 0          & 1          \\ \cline{2-2} \cline{9-14} 
                       & INTT & \multicolumn{6}{c|}{}                                                                                                                                                                  & 1          & 1          & 1          & 0          & 1          & 0          \\ \cline{2-2} \cline{9-14} 
                       & PWM0 & \multicolumn{6}{c|}{}                                                                                                                                                                  & 0          & 0          & 1          & 1          & 0          & 0          \\ \cline{2-2} \cline{9-14} 
                       & PWM1 & \multicolumn{6}{c|}{}                                                                                                                                                                  & 0          & 1          & 1          & 1          & 0          & 0          \\ \hline
\end{tabular}
\caption{Control Signals of Kyber BFUs during NTT/NTT$^{-1}$/PWM}
\label{tab:kyberbfucontrol}
\end{table}
CRYSTALS-Kyber does not support radix-4 NTT because of its property of incomplete NTT. The authors in~\cite{duong2022configurable} has proposed a mixed-radix implementation of NTT multiplication for Kyber with a very high area overhead $3918$ LUTs and $26$ DSPs.
\vspace{-0.1 in}
\subsection{Polynomial Multiplication Unit for Dilithium}
\vspace{-0.05 in}
\label{reddilithium}
CRYSTALS-Dilithium has a 23-bit modulus with structure $2^{23}-2^{13}+1$. 
We have chosen Montgomery modular multiplication for Dilithium using the prime structure with an overhead of $227$ LUTs and $2$ DSPs for one $23 \times 23$ bits multiplication.
Our NTT multiplication unit for Dilithum processes $2$ Dilithium coefficients at every clock cycle using $1$ BFU. We have used two simple dual-port RAMs of size $256*24$ for coefficient storage along with one $18K$ single port ROM for twiddle factors and one $36K$ single port ROM for address ROM. Modular addition/subtraction for Dilithium is the same as that of Kyber with prime value $8380417$.
The butterfly unit for CRYSTALS-Dilithium has been configured similarly to Kyber to perform NTT/NTT$^{-1}$/PWM depending upon select lines of different multiplexers. As PWM is much simpler in Dilithium, the complexity of the BFU structure reduces. The NTT polynomial unit for Dilithium with $1$ BFU has a latency of $2304$ with an overhead of $724$ LUTs, $769$ FFs, $2$ DSPs, $2.5$ BRAMs with a frequency of $413$ MHz on Zynq Ultrascale+ FPGA.

\begin{table*}[]
\centering
\scriptsize
\begin{tabular}{|c|c|c|c|c|c|c|ccc|c|ccc|c|c|}
\hline
\textbf{NTT Type}          & \textbf{}                    & \textbf{Board}           & \textbf{LUT} & \textbf{FFs} & \textbf{BRAM}                 & \textbf{DSP}                & \multicolumn{3}{c|}{\textbf{Latency}}                                                  & \textbf{\begin{tabular}[c]{@{}c@{}}Freq.\\ (Mhz)\end{tabular}} & \multicolumn{3}{c|}{\textbf{ADP-LUT}}                                                  & \textbf{\begin{tabular}[c]{@{}c@{}}ADP-\\ DSP\end{tabular}} & \textbf{\begin{tabular}[c]{@{}c@{}}ADP-\\ BRAM\end{tabular}} \\ \hline
                           &                              &                          &              &              &                               &                             & \multicolumn{1}{c|}{\textbf{NTT}}  & \multicolumn{1}{c|}{\textbf{INTT}} & \textbf{PWM} & \textbf{}                                                      & \multicolumn{1}{c|}{\textbf{NTT}}  & \multicolumn{1}{c|}{\textbf{INTT}} & \textbf{PWM} &                                                             &                                                              \\ \hline
\multirow{8}{*}{Kyber}     & \cite{bisheh2021high}                            & \multirow{7}{*}{Artix-7} & 801          & 717          & 2                             & 4                           & \multicolumn{1}{c|}{324}           & \multicolumn{1}{c|}{324}           & -            & 222                                                            & \multicolumn{1}{c|}{1169}          & \multicolumn{1}{c|}{1169}          & {-}     & 6                                                           & 3                                                            \\ \cline{2-2} \cline{4-16} 
                           & \cite{bisheh2021instruction}                           &                          & 360          & 145          & 2                             & 3                           & \multicolumn{1}{c|}{940}           & \multicolumn{1}{c|}{1203}          & 1289         & 115                                                            & \multicolumn{1}{c|}{2943}          & \multicolumn{1}{c|}{3766}          & 4035         & 25                                                          & 16                                                           \\ \cline{2-2} \cline{4-16} 
                           & \cite{bisheh2021instruction}                           &                          & 737          & 290          & 4                             & 6                           & \multicolumn{1}{c|}{474}           & \multicolumn{1}{c|}{602}           & 1289         & 115                                                            & \multicolumn{1}{c|}{3038}          & \multicolumn{1}{c|}{3858}          & 8261         & 25                                                          & 16                                                           \\ \cline{2-2} \cline{4-16} 
                           & \cite{xing2021compact}                            &                          & 1579         & 1058         & 3                             & 2                           & \multicolumn{1}{c|}{448}           & \multicolumn{1}{c|}{448}           & 256          & 161                                                            & \multicolumn{1}{c|}{4394}          & \multicolumn{1}{c|}{4394}          & 2511         & 6                                                           & 8                                                            \\ \cline{2-2} \cline{4-16} 
                           & \cite{dang2022high}                            &                          & 880          & 999          & 1.5                           & 2                           & \multicolumn{1}{c|}{448}           & \multicolumn{1}{c|}{448}           & 256          & 222                                                            & \multicolumn{1}{c|}{1776}          & \multicolumn{1}{c|}{1776}          & 1015         & 4                                                           & 3                                                            \\ \cline{2-2} \cline{4-16} 
                           & \cite{zhang2021towards}                        &                          & 609          & 640          & 4                             & 2                           & \multicolumn{1}{c|}{490}           & \multicolumn{1}{c|}{490}           & -            & 256                                                            & \multicolumn{1}{c|}{1166}          & \multicolumn{1}{c|}{1166}          & {-}     & 8                                                           & 15                                                           \\ \cline{2-2} \cline{4-16} 
                           & \textbf{TW}                  &                          & \textbf{799} & \textbf{916} & \multirow{2}{*}{\textbf{2}}   & \multirow{2}{*}{\textbf{2}} & \multicolumn{1}{c|}{\textbf{448}}  & \multicolumn{1}{c|}{\textbf{448}}  & \textbf{256} & \textbf{310}                                                   & \multicolumn{1}{c|}{\textbf{1155}} & \multicolumn{1}{c|}{\textbf{1155}} & \textbf{660} & \textbf{3}                                                  & \textbf{3}                                                   \\ \cline{2-5} \cline{8-16} 
                           & \textbf{TW}                  & ZCU+ 102                 & \textbf{698} & \textbf{865} &                               &                             & \multicolumn{1}{c|}{\textbf{448}}  & \multicolumn{1}{c|}{\textbf{448}}  & \textbf{256} & \textbf{500}                                                   & \multicolumn{1}{c|}{\textbf{625}}  & \multicolumn{1}{c|}{\textbf{625}}  & \textbf{357} & \textbf{2}                                                  & \textbf{2}                                                   \\ \hline
\multirow{8}{*}{Dilithium} & \cite{beckwith2021high}                            & Artix-7                  & 9018         & 6292         & 2                             & 16                          & \multicolumn{1}{c|}{256}           & \multicolumn{1}{c|}{256}           & -            & 250                                                            & \multicolumn{1}{c|}{9234}          & \multicolumn{1}{c|}{9234}          & {-}     & 16                                                          & 2                                                            \\ \cline{2-16} 
                           & \cite{wang2022efficient}                            & Zynq 7000                & 2386         & 932          & 2                             & 8                           & \multicolumn{1}{c|}{256}           & \multicolumn{1}{c|}{256}           & 64           & 217                                                            & \multicolumn{1}{c|}{2815}          & \multicolumn{1}{c|}{2815}          & 704          & 9                                                           & 2                                                            \\ \cline{2-16} 
                           &\cite{zhao2022compact}                              & \multirow{2}{*}{Artix-7} & 524          & 759          & 1                             & 17                          & \multicolumn{1}{c|}{533}           & \multicolumn{1}{c|}{536}           & -            & 311                                                            & \multicolumn{1}{c|}{898}           & \multicolumn{1}{c|}{903}           & {-}     & 29                                                          & 2                                                            \\ \cline{2-2} \cline{4-16} 
                           & \cite{gupta2023lightweight}                         &                          & 2759         & 2037         & 7                             & 4                           & \multicolumn{1}{c|}{512*}           & \multicolumn{1}{c|}{512*}           & 128*          & 163                                                            & \multicolumn{1}{c|}{8666}          & \multicolumn{1}{c|}{8666}          & 2167         & 13                                                          & 22                                                           \\ \cline{2-16} 
                           & \cite{gupta2023lightweight}                         & ZCU+                     & 2759         & 2037         & 7                             & 4                           & \multicolumn{1}{c|}{512*}           & \multicolumn{1}{c|}{512*}           & 128*          & 391                                                            & \multicolumn{1}{c|}{3613}          & \multicolumn{1}{c|}{3613}          & 903          & 5                                                           & 9                                                            \\ \cline{2-16} 
                           & \multirow{3}{*}{\textbf{TW}} & Zynq 7000                & \textbf{698} & \textbf{771} & \multirow{3}{*}{\textbf{2.5}} & \multirow{3}{*}{\textbf{2}} & \multicolumn{1}{c|}{\textbf{1024}} & \multicolumn{1}{c|}{\textbf{1024}} & \textbf{256} & \textbf{279}                                                   & \multicolumn{1}{c|}{\textbf{2562}} & \multicolumn{1}{c|}{\textbf{2562}} & \textbf{640} & \textbf{7}                                                  & \textbf{9}                                                   \\ \cline{3-5} \cline{8-16} 
                           &                              & Artix-7                  & \textbf{690} & \textbf{771} &                               &                             & \multicolumn{1}{c|}{\textbf{1024}} & \multicolumn{1}{c|}{\textbf{1024}} & \textbf{256} & \textbf{273}                                                   & \multicolumn{1}{c|}{\textbf{2588}} & \multicolumn{1}{c|}{\textbf{2588}} & \textbf{647} & \textbf{8}                                                  & \textbf{9}                                                   \\ \cline{3-5} \cline{8-16} 
                           &                              & ZCU+ 102                 & \textbf{724} & \textbf{769} &                               &                             & \multicolumn{1}{c|}{\textbf{1024}} & \multicolumn{1}{c|}{\textbf{1024}} & \textbf{256} & \textbf{413}                                                   & \multicolumn{1}{c|}{\textbf{1795}} & \multicolumn{1}{c|}{\textbf{1795}} & \textbf{449} & \textbf{5}                                                  & \textbf{6}                                                   \\ \hline
\end{tabular}
\caption{Comparison Table for Kyber/Dilithium NTT Multiplication unit. * marked results are estimated.}
\label{tab:compareDesigns1}
\end{table*}
\begin{table*}[]
\centering
\scriptsize
\begin{tabular}{|c|c|c|c|c|c|c|c|ccc|c|ccc|c|c|}
\hline
\textbf{NTT Type}         & \textbf{}                     & \textbf{Board}            & \textbf{LUT}                   & \textbf{FFs}                   & \textbf{BRAM}                 & \textbf{DSP}                & \textbf{K/D} & \multicolumn{3}{c|}{\textbf{Latency}}                                                  & \textbf{\begin{tabular}[c]{@{}c@{}}Freq.\\ (Mhz)\end{tabular}} & \multicolumn{3}{c|}{\textbf{ADP-LUT}}                                                   & \textbf{\begin{tabular}[c]{@{}c@{}}ADP-\\ DSP\end{tabular}} & \textbf{\begin{tabular}[c]{@{}c@{}}ADP-\\ BRAM\end{tabular}} \\ \hline
                          &                               &                           &                                &                                &                               &                             & \textbf{}    & \multicolumn{1}{c|}{\textbf{NTT}}  & \multicolumn{1}{c|}{\textbf{INTT}} & \textbf{PWM} & \textbf{}                                                      & \multicolumn{1}{c|}{\textbf{NTT}}  & \multicolumn{1}{c|}{\textbf{INTT}} & \textbf{PWM}  &                                                             &                                                              \\ \hline
\multirow{14}{*}{Unified} & \multirow{2}{*}{\cite{aikata2022kali}}         & \multirow{8}{*}{ZCU+ 102} & \multirow{2}{*}{3487}          & \multirow{2}{*}{1918}          & \multirow{2}{*}{3*}            & \multirow{2}{*}{4}          & K            & \multicolumn{1}{c|}{224}           & \multicolumn{1}{c|}{224}           & 128          & 270                                                            & \multicolumn{1}{c|}{2893}          & \multicolumn{1}{c|}{2893}          & 1653          & 3                                                           & 2                                                            \\ \cline{8-17} 
                          &                               &                           &                                &                                &                               &                             & D            & \multicolumn{1}{c|}{512}           & \multicolumn{1}{c|}{512}           & 128          & 270                                                            & \multicolumn{1}{c|}{6612}          & \multicolumn{1}{c|}{6612}          & 1653          & 8                                                           & 6                                                            \\ \cline{2-2} \cline{4-17} 
                          & \multirow{12}{*}{\textbf{TW}} &                           & \multirow{2}{*}{\textbf{1384}} & \multirow{2}{*}{\textbf{1220}} & \multirow{2}{*}{\textbf{4.5}} & \multirow{2}{*}{\textbf{2}} & \textbf{K}   & \multicolumn{1}{c|}{\textbf{448}}  & \multicolumn{1}{c|}{\textbf{448}}  & \textbf{256} & \textbf{387}                                                   & \multicolumn{1}{c|}{\textbf{1602}} & \multicolumn{1}{c|}{\textbf{1602}} & \textbf{916}  & \textbf{2}                                                  & \textbf{5}                                                   \\ \cline{8-17} 
                          &                               &                           &                                &                                &                               &                             & \textbf{D}   & \multicolumn{1}{c|}{\textbf{1024}} & \multicolumn{1}{c|}{\textbf{1024}} & \textbf{256} & \textbf{387}                                                   & \multicolumn{1}{c|}{\textbf{3662}} & \multicolumn{1}{c|}{\textbf{3662}} & \textbf{916}  & \textbf{5}                                                  & \textbf{12}                                                  \\ \cline{4-17} 
                          &                               &                           & \multirow{2}{*}{\textbf{2893}} & \multirow{2}{*}{\textbf{2356}} & \multirow{2}{*}{\textbf{4.5}} & \multirow{2}{*}{\textbf{4}} & \textbf{K}   & \multicolumn{1}{c|}{\textbf{224}}  & \multicolumn{1}{c|}{\textbf{224}}  & \textbf{128} & \textbf{342}                                                   & \multicolumn{1}{c|}{\textbf{1895}} & \multicolumn{1}{c|}{\textbf{1895}} & \textbf{1083} & \textbf{3}                                                  & \textbf{3}                                                   \\ \cline{8-17} 
                          &                               &                           &                                &                                &                               &                             & \textbf{D}   & \multicolumn{1}{c|}{\textbf{512}}  & \multicolumn{1}{c|}{\textbf{512}}  & \textbf{128} & \textbf{342}                                                   & \multicolumn{1}{c|}{\textbf{4331}} & \multicolumn{1}{c|}{\textbf{4331}} & \textbf{1083} & \textbf{6}                                                  & \textbf{7}                                                   \\ \cline{4-17} 
                          &                               &                           & \multirow{2}{*}{\textbf{5909}} & \multirow{2}{*}{\textbf{3376}} & \multirow{2}{*}{\textbf{5.5}} & \multirow{2}{*}{\textbf{8}} & \textbf{K}   & \multicolumn{1}{c|}{\textbf{112}}  & \multicolumn{1}{c|}{\textbf{112}}  & \textbf{64}  & \textbf{294}                                                   & \multicolumn{1}{c|}{\textbf{2251}} & \multicolumn{1}{c|}{\textbf{2251}} & \textbf{1286} & \textbf{3}                                                  & \textbf{2}                                                   \\ \cline{8-17} 
                          &                               &                           &                                &                                &                               &                             & \textbf{D}   & \multicolumn{1}{c|}{\textbf{256}}  & \multicolumn{1}{c|}{\textbf{256}}  & \textbf{64}  & \textbf{294}                                                   & \multicolumn{1}{c|}{\textbf{5145}} & \multicolumn{1}{c|}{\textbf{5145}} & \textbf{1286} & \textbf{7}                                                  & \textbf{5}                                                   \\ \cline{3-17} 
                          &                               & \multirow{6}{*}{Artix-7}  & \multirow{2}{*}{\textbf{1315}} & \multirow{2}{*}{\textbf{1280}} & \multirow{2}{*}{\textbf{4.5}} & \multirow{2}{*}{\textbf{2}} & \textbf{K}   & \multicolumn{1}{c|}{\textbf{448}}  & \multicolumn{1}{c|}{\textbf{448}}  & \textbf{256} & \textbf{263}                                                   & \multicolumn{1}{c|}{\textbf{2240}} & \multicolumn{1}{c|}{\textbf{2240}} & \textbf{1280} & \textbf{3}                                                  & \textbf{8}                                                   \\ \cline{8-17} 
                          &                               &                           &                                &                                &                               &                             & \textbf{D}   & \multicolumn{1}{c|}{\textbf{1024}} & \multicolumn{1}{c|}{\textbf{1024}} & \textbf{256} & \textbf{263}                                                   & \multicolumn{1}{c|}{\textbf{5120}} & \multicolumn{1}{c|}{\textbf{5120}} & \textbf{1280} & \textbf{8}                                                  & \textbf{18}                                                  \\ \cline{4-17} 
                          &                               &                           & \multirow{2}{*}{\textbf{3105}} & \multirow{2}{*}{\textbf{2389}} & \multirow{2}{*}{\textbf{4.5}} & \multirow{2}{*}{\textbf{4}} & \textbf{K}   & \multicolumn{1}{c|}{\textbf{224}}  & \multicolumn{1}{c|}{\textbf{224}}  & \textbf{128} & \textbf{200}                                                   & \multicolumn{1}{c|}{\textbf{3478}} & \multicolumn{1}{c|}{\textbf{3478}} & \textbf{1987} & \textbf{4}                                                  & \textbf{5}                                                   \\ \cline{8-17} 
                          &                               &                           &                                &                                &                               &                             & \textbf{D}   & \multicolumn{1}{c|}{\textbf{512}}  & \multicolumn{1}{c|}{\textbf{512}}  & \textbf{128} & \textbf{200}                                                   & \multicolumn{1}{c|}{\textbf{7949}} & \multicolumn{1}{c|}{\textbf{7949}} & \textbf{1987} & \textbf{10}                                                 & \textbf{12}                                                  \\ \cline{4-17} 
                          &                               &                           & \multirow{2}{*}{\textbf{6201}} & \multirow{2}{*}{\textbf{3562}} & \multirow{2}{*}{\textbf{5.5}} & \multirow{2}{*}{\textbf{8}} & \textbf{K}   & \multicolumn{1}{c|}{\textbf{112}}  & \multicolumn{1}{c|}{\textbf{112}}  & \textbf{64}  & \textbf{165}                                                   & \multicolumn{1}{c|}{\textbf{4209}} & \multicolumn{1}{c|}{\textbf{4209}} & \textbf{2405} & \textbf{5}                                                  & \textbf{4}                                                   \\ \cline{8-17} 
                          &                               &                           &                                &                                &                               &                             & \textbf{D}   & \multicolumn{1}{c|}{\textbf{256}}  & \multicolumn{1}{c|}{\textbf{256}}  & \textbf{64}  & \textbf{165}                                                   & \multicolumn{1}{c|}{\textbf{9621}} & \multicolumn{1}{c|}{\textbf{9621}} & \textbf{2405} & \textbf{12}                                                 & \textbf{9}                                                   \\ \hline
\end{tabular}
\caption{Comparison Table for Unified NTT Multiplication Unit. * marked results are estimated.}
\label{tab:compareDesigns2}
\vspace{-0.2 in}
\end{table*}
\begin{figure}[H]
    \centering
    \includegraphics[clip,scale=0.3]{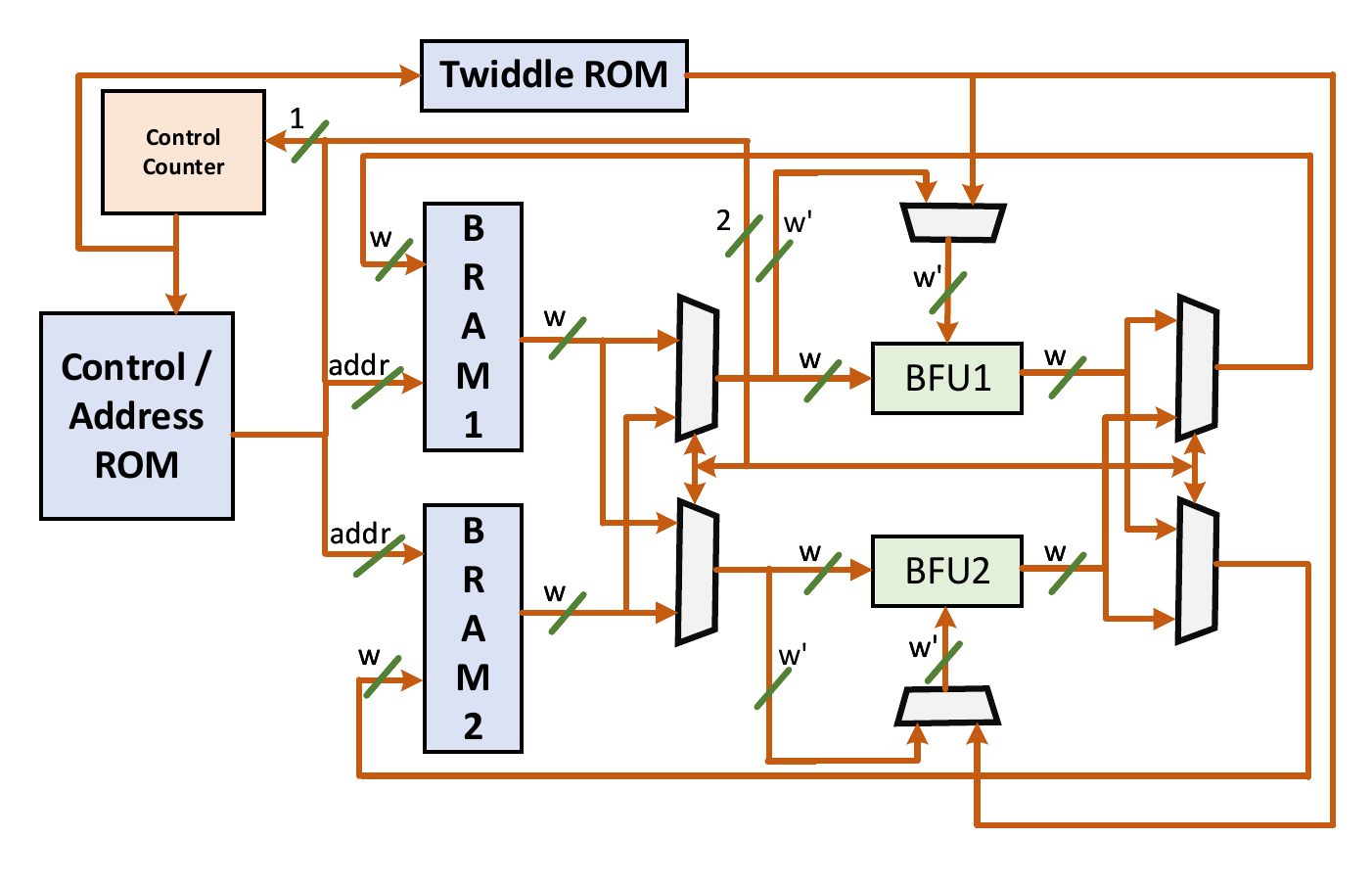}
    \caption{Unified NTT Core Architecture (BRAM 1, BRAM 2: Coefficient Storage. BRAM 3: Read Write Address for BRAM 1 and BRAM 2 and Read/Write Control. Mem 4: Twiddle Storage.))}
    \label{fig:nttcore}
\end{figure}
\vspace{-0.2 in}
\section{Unified NTT Multiplication Unit}
\vspace{-0.1 in}
\label{sec:unified}
Now, we will focus on the unified NTT multiplication for CRYSTALS-Kyber and CRYSTALS-Dilithium. Although for CRYSTALS-Dilithium, radix-4 NTT is an efficient choice over radix-2 NTT, radix-4 NTT is not suitable for Kyber as discussed in section~\ref{polykyber}. 
So, radix-2 NTT  is our choice for unified NTT architecture for Kyber and Dilithium. We have started by supporting $2$ Kyber BFUs that can also be used as $1$ Dilithium BFU. Thus, we can process $4$ coefficients of Kyber or $2$ coefficients of Dilithium simultaneously.
\subsection{Unified NTT Core with 2 BFUs of Kyber and 1BFU for Dilithium}
\label{kd21}
We have used two DSP blocks of size $23*12$ bit to support the $2$ multiplication of Kyber/1 multiplication in Dilithium. For Kyber, the output of the multiplications is fed directly to the modular reduction unit, whereas for Dilithium, the partial products are added and then forwarded to the modular reduction unit of Dilithium. An architectural diagram of our Unified NTT Core architecture is shown in Fig.\ref{fig:nttcore}. In Fig. \ref{fig:nttcore}, we have used $2$ BFUs of Kyber, which can be configured as $1$ BFU of Dilithium. Each BRAM produces $24$ bit data for the two BFUs of Kyber that consume $4$ coefficients in every clock cycle. For Dilithium, the same data can drive $1$ BFU which consumes two coefficients of Dilithium in every clock cycle. This is possible as the coefficient size of Dilithium is nearly double of Kyber's coefficient size.
The read/write addresses are stored according to Algorithm. \ref{algorithm1}. The control counter also controls the access of twiddle ROM, where all the twiddle factors of Kyber and Dilithium are stored.

The new reconfigurable BFU unit supports NTT/INTT/PWM for CRYSTALS-Kyber and CRYSTALS-Dilithium. The latency of both NTT and NTT$^{-1}$ for Kyber is $448$ cycles with $2$ BFUs, and for Dilithium, the latency is $1024$ with $1$ BFU. The pipeline depth of the design is $15$. The latency for PWM of Kyber and Dilithium is $256$ cycles. The architectural diagram of the BFU is similar to Fig. \ref{fig:bfukyber2}. Apart from DSPs, the proposed reconfigurable BFUs also uses shared adder and subtractors. 
\begin{figure}[h]
    \centering
    \vspace{-0.2 in}
    \includegraphics[clip,scale=0.4]{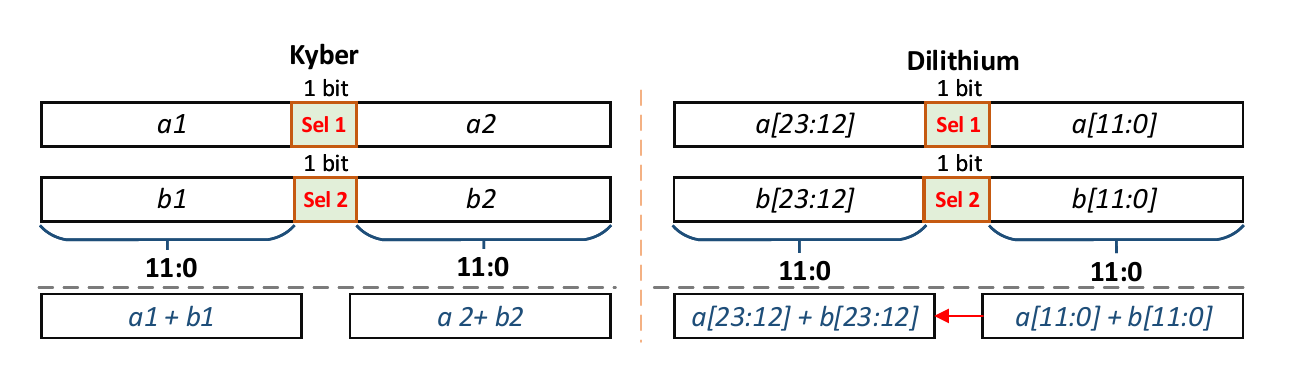}
    \caption{Shared adder/subtractor of Kyber/Dilithium}
    \label{fig:addsub}
    \vspace{-0.1 in}
\end{figure}
The diagram of shared adder/subtractor is given in Fig. \ref{fig:addsub}. To utilize the fast carry chain inside FPGAs, an extra bit is inserted in the $12^{th}$ position of the coefficients, depending upon the addition/subtraction operation for Kyber/Dilithium. The selection bits $sel1/sel2$ will decide if the carry/borrow propagation will happen or not. In the case of Kyber, the $12^{th}$ bit will be used stop the carry propagation to the next addition, whereas, for Dilithium, the $12^{th}$ bit will forward the carry/borrow to the next half of addition/subtraction. Thus with this module, we can perform two addition/subtractions for Kyber and one addition/subtraction for Dilithium. The design of~\cite{aikata2022kali} uses additional multiplexers to achieve this and hence can not take advantage of the fast carry chain of FPGAs.

The four configurations for addition/subtraction are given below:
\begin{compactitem}
    \item Subtraction of Kyber/addition of Dilithium: $sel1=1$ , $sel2=0$.
    \item Subtraction of Dilithium/addition of Kyber: $sel1=0$, $sel2=0$.
\end{compactitem}
This unified architecture consumes $1384$ LUTs, $4.5$ BRAMs, $2$ DSPs with a frequency of $387 MHz$ in the Zynq Ultrascale+ FPGA.
\vspace{-0.05 in}
\subsection{Unified NTT with 4 BFUs of Kyber and 2 BFUs for Dilithium}
To fit $4$ BFUs of CRYSTALS-Kyber and $2$ BFUs of CRYSTALS-Dilithium, we need to access $8$ coefficients of Kyber and $4$ coefficients of Dilithium at a time. In this case, our coefficient RAMs are of size $48*128$. The address ROM occupies one $18K$ ROM and a $36K$ ROM along with the twiddle ROM of size $36K$. We kept the pipeline depth the same as in section~\ref{kd21}. NTT and NTT$^{-1}$ operation of Kyber consumes $224$ cycles each and a latency of $128$ cycles is required for PWM. NTTa nd NTT$^{-1}$ of Dilithium completes in $512$ cycles, each along with a latency of $128$ cycles for PWM. This unified architecture consumes $2893$ LUTs, $4.5$ BRAMs and $4$ DSPs with a frequency of $342$ MHz in Zynq Ultrascale+ FPGA.
\vspace{-0.05 in}
\subsection{Unified NTT Core with 8 BFUs of Kyber and 4 BFUs for Dilithium}
Eight BFUs of Kyber will consume $16$ coefficients of Kyber and $8$ coefficients of Dilithium at a time. So, we need two simple dual-port RAMs of size $96*64$ that consume one $36K$ BRAM and one $18K$ BRAM each. The address ROM consumes a BRAM of size $36K$, and the twiddle ROM consumes one $36K$ BRAM with one $18K$ BRAM. For the previous two implementations, we kept the pipeline cycle at $15$. But in the case of $8$ radix-2 access in parallel for Kyber, the pipelining depth can not be more than $8$ as explained in section~\ref{sec:conflictfree}. 
 As a result, a slight decrease in the frequency can be observed. This unified architecture consumes $5909$ LUTs, $5.5$ BRAMs and $8$ DSPs and achieved a frequency of $294$ MHz in Zynq Ultrascale+ FPGA.
\vspace{-0.05 in}
\section{Comparision Results and Analysis}
\vspace{-0.05 in}
\label{sec:experiment}
In Tab.~\ref{tab:compareDesigns1}, we show the comparison between our proposed standalone implementation of Kyber and Dilithium's NTT architecture, whereas Tab.~\ref{tab:compareDesigns2} shows the comparative analysis for our proposed unified NTT design.
We have used Artix-7 (XC7A200TFBG676-3), ZCU+ 102 (XCZU9EGFFV-B11562e) and Zynq-7000 (XCZU9EG-FFVB1156-2-e) FPGA as our target platform. ADP-LUT, ADP-DSP and ADP-BRAM are chosen as the performance metrics, calculated by the formulas $\frac{latency}{frequency} * LUTs$, $\frac{latency}{frequency} * DSPs$ and $\frac{latency}{frequency} * BRAMs$ respectively. As, some of the works have not reported latency cycles for PWM operation, for computing ADP-DSP or ADP-BRAM, we have used the latency cycles for NTT only. Apart from~\cite{bisheh2021instruction,land2021hard}, all the state-of-the-art designs have the same latency for NTT and NTT$^{-1}$. 
We have not added \cite{li2022reconfigurable} to our comparison table as their modular adder/subtractor is prone to produce erroneous values, as mentioned in section \ref{polykyber}. 
We have also not added \cite{zhao2022compact} to our comparison table as the design has not reported the latency cycles for NTT/NTT$^{-1}$/PWM. Also, this design achieved a very low frequency ($97$ MHz only) compared to other existing works. The result in Tab.~\ref{tab:compareDesigns1} clearly shows that our standalone implementation of NTT architecture for Kyber is not only the most compact but also achieves superior ADP-LUT, ADP-DSP and ADP-BRAM values. 
The ATP-LUT product for NTT multiplication unit of Dilithium in \cite{land2021hard}, is superior to our achieved ATP-LUT. But ATP-DSP of our design is four times less than that of \cite{land2021hard}. 

For the unified Kyber/Dilithium, the ADP product of all of our designs in ZCU+ shows better results than~\cite{aikata2022kali}as shown in Tab.~\ref{tab:compareDesigns2}. Authors in \cite{aikata2022kali}, proposed a unified NTT Core Architecture that supports $4$ BFUs of Kyber as well as $2$ BFUs of Dilithium with $3487$ LUTs,$4$ DSPs. The authors have not provided their BRAM consumption for the NTT multiplication. We have estimated it to be $3$ (two for coefficient storage and one for storing twiddle factors). In Table~\ref{tab:compareDesigns2}, we have used the symbols $\mathbf{K}$ and $\mathbf{D}$ to indicate the latency of Kyber and Dilithium's NTT operation.
The authors used a combined Solinas prime reduction methodology for Kyber/Dilithium. We discovered that our proposed Montgomery multiplication, which takes advantage of the structure of the quotients, consumes less overhead than the Solinas prime reduction. This results in the superior performance of our proposed architecture. Even our unified design 1 (2 radix-2 BFUs for
Kyber that can also be used as 1 radix-2 BFU of Dilithium) 
achieves better result than~\cite{aikata2022kali}, tested on ZCU+ 102 FPGA.
\vspace{-0.12 in}
\section{Conclusion}\label{sec:conc}
\vspace{-0.1 in}
In this paper, we have explored the unified radix-2 NTT architecture to its maximum efficiency. Our standalone implementations of NTT multiplier for Kyber and Dilithium show excellent results when compared with existing implementations. Additionally, our proposed unified implementations consume lower area overhead with a significant improvement in frequency compared to existing unified architecture. 
This work can be used to develop a unified architecture for a post-quantum secure public-key framework. 
\vspace{-0.1 in}
\bibliographystyle{IEEEtran}
\bibliography{ref}

\end{document}